\documentclass[11pt]{JHEP3}
\usepackage{amsmath, amssymb, bm,cite,axodraw}

\title{BMN operators with a scalar fermion pair and operator mixing in ${\cal N}=4$ Super Yang-Mills Theory}
\author{ 
  Zhiguang~Xiao\\

  School of Physics and Astronomy,  University of Southampton\\
  Highfield, Southampton, SO17 1BJ, U.K.\\

  E-mail: 
\email{z.g.xiao@phys.soton.ac.uk
} 
}

\preprint{}


\abstract{The mixings between BMN operators with two scalar impurities
and those with a scalar fermion pair are discussed to the lowest order
at planar level. For this purpose, matrix model effective vertices are
calculated to $O(g^3)$. All the mixing patterns are explicitly
obtained.  }

\begin{document}
\section{Introduction}
${\cal N}=4$ supersymmetric Yang-Mills theory attracts a lot of
interest after the discovery of AdS/CFT correspondence.  During the
early years, due to the BPS properties of the states from dimensional reduction
of the supergravity, studies  on the CFT side were mainly focused on
the protected operators (see \cite{Aharony:1999ti} for a review).  The
proposal of BMN correspondence \cite{Berenstein:2002jq} provides
further evidence of AdS/CFT beyond supergravity.  Operators ${\rm
Tr}(Z^{J})$ are BPS and correspond to the string vacuum states
$|0,p^+\rangle$ in the light-cone quantization of the string in a pp
wave background, whereas operators like ${\rm Tr}(\phi_i Z^p\phi_j
Z^{J-p})$ are not protected and get mixed in order to remain the
eigenstates of the one-loop dilatation operator. In the BMN limit
these eigenstate operators (BMN operators) correspond to the string
excitation states
$\alpha_n^{i\dagger}\alpha_n^{j\dagger}|0,p^+\rangle$ (see e.g.
\cite{Plefka:2003nb} for a pedagogical review).  BMN operators
involving vector \cite{Gursoy:2002yy,Klose:2003tw,Chu:2003ji} and
fermions
\cite{Eden:2003sj,Bianchi:2003eg,Georgiou:2004ty,Dobashi:2006fu} are
also examined.  Exact BMN operators at finite $N$ and $J$ are obtained
in \cite{Beisert:2002tn} and extended to the whole superconformal
multiplet incorporating BMN operators with various combinations of two
impurities by using superconformal symmetry. Operator mixings beyond
one loop in the scalar sector are also discussed in
\cite{Beisert:2003tq}. 

Unlike the mixings in the pure scalar sector, mixings between scalar and
fermionic sectors are less studied. In \cite{Eden:2003sj}, some low
dimension operator mixings are discussed. Instanton contributions to
their mixings are also studied in \cite{Green:2005ib}.  After mapping
the operators into spin chains and the dilatation operator into
a spin-chain Hamiltonion, a Bethe ansatz can be used to solve the  mixing
problem \cite{Minahan:2002ve,Staudacher:2004tk}.  Integrability and the
analytic Bethe ansatz make it possible to obtain anomalous dimensions
for operators with more than two impurities.  For ${\cal N}=4$ SUSY YM, this
was first proposed in \cite{Minahan:2002ve} for an $so(6)$ spin chain and
further extended to the $psu(2,2|4)$ super
spin chain in \cite{Beisert:2003yb}. A subsector $su(2|3)$ dynamic spin
chain which includes the mixing between fermion pairs and bosons is
discussed along these lines in \cite{Beisert:2003ys}. The coordinate-space
Bethe ansatz in calculating the spin-chain S-matrix can also produce
the eigenstates with multiple 
impurities \cite{Staudacher:2004tk,Beisert:2005fw}. This is known as  
the perturbative asymptotic Bethe ansatz in this context. The $su(2|2)$
dynamic S-matrix of the $su(2|3)$ subsector  was subsequently
discussed using the nested Bethe ansatz \cite{Beisert:2005tm}. In this
subsector, the fermions form singlets of the $SU(3)$ subgroup of the R
symmetry. 
 
In the present paper, the fermions we will discuss transform under the
$SO(4)=SU(2)\times SU(2)$ subgroup of the $SU(4)_R$ symmetry as two
fundamental representations for the two $SU(2)$'s respectively. As a first
attempt, we only discuss the mixings between BMN operators with two
impurities at planar level and the fermion pair form a Lorentz scalar.
In particular, we will discuss the mixings between operators:
\begin{eqnarray}
{\cal O}^{-1,J}_{1}&=&\left(\frac{2\pi}{\sqrt N}\right)^{J+2}{\rm Tr}(\bar Z Z^{J+1})
\\
{\cal O}^{p,J}_{ij}&=&\left(\frac{2\pi}{\sqrt
N}\right)^{J+2}{\rm
Tr}(\phi_i Z^p \phi_j Z^{j-p}) \,,\quad \text{ for } i,j=1,2,4,5
\label{eq:O-ij}\\
\Psi^{p,J}_{ij}&=&\left(\frac{2\pi}{\sqrt
N}\right)^{J}\left(\frac{2^{1/4}\pi}{\sqrt N}\right)^{2}{\rm
Tr}(\psi^{i,\alpha} Z^p \psi^j_{\alpha} Z^{J-p})\,,\quad \text{ for }
i,j=1,2\,,
\label{eq:Psi-ij}\\
\Psi^{p,J}_{ij}&=&\left(\frac{2\pi}{\sqrt
N}\right)^{J}\left(\frac{2^{1/4}\pi}{\sqrt N}\right)^{2}{\rm Tr}(\bar\psi_{i,\dot \alpha} Z^p
\bar\psi^{\dot\alpha}_j Z^{J-p})\,,\quad \text{ for } i,j=3,4\,.
\label{eq:operators}
\end{eqnarray}
We have normalized the operators so that 
\begin{eqnarray}
\langle {\cal O}_{ij}^{p,J}(x)\bar{\cal O}_{ij}^{p,J}(0)\rangle=\frac
1{x^{2(J+2)}}\,,
\nonumber \\
\langle \Psi_{ij}^{p,J-1}(x)\bar\Psi_{ij}^{p,J-1}(0)\rangle=\frac
1{x^{2(J+2)}}
\end{eqnarray}
for tree level planar diagrams. From the $SO(4)$ point of view, since
scalars with two impurities can form single trace operators of
$1+3+\bar 3+9$ representations, operators like (\ref{eq:Psi-ij}) with
$i=1, 2$ and $j=3,4$ which is in vector $4$ representation can not be
mixed with (\ref{eq:O-ij}). Because the common representations of two
scalars or two spinors are $1+3+\bar 3$, we expect that they only mix
in these representations.

In \cite{Georgiou:2008vk}, a method is developed to obtain the mixing
between these operators using the property that the eigenstate
operator should be annihilated by some supercharges. We will reproduce
their results by directly diagonalizing the dilatation operator and give
all the mixing patterns. Since we only consider two impurity cases, it
is easy to do it without using Bethe ansatz. We will follow the method
used in
\cite{Constable:2002hw,Beisert:2002bb,Constable:2002vq,Kristjansen:2002bb,Beisert:2002ff}
to deduce the dilatation operator in the name of effective vertices.
Even though the algebraic method was used to extend the scalar one-loop
dilatation operator to the full ${\cal N}=4$ SUSY YM
\cite{Beisert:2003jj}, we will still deduce the one-loop effective
vertices by feynman diagrams since it is not hard to calculate the
correlation functions with only scalar fermion pairs
\cite{Georgiou:2004ty,Dobashi:2006fu} and we can formulate the
effective vertices taking advantage of the scalar property of the
fermion pair. We also need to calculate the effective vertices to
$O(g^3)$ which is the lowest order at which the mixing between scalar
sector and fermionic sector occurs.

In section \ref{sect:Matrix-oneloop}, we review the matrix model
effective vertices and their relation with dilatation operators and
calculate the one-loop effective vertices with one scalar fermion
pair. In section \ref{sect:Mixing-oneloop}, we review the mixing of
pure bosonic operators using the one-loop effective vertices and
obtain the mixing of operators with one scalar fermion pair using the
results from pure scalar operator mixing. In section \ref{sect:Order},
we perform a general order analysis about the form of the eigenstates.
In section \ref{sect:Matrix-O3}, we give the matrix model vertices to
$O(g^3)$ with a pair of fermions.  In section \ref{sect:Mixing-O3}, we
use the effective vertices to obtain the dilatation matrix to $O(g^3)$
and solve the eigenvector operators.  The results are consistent with
\cite{Georgiou:2008vk}. Section \ref{sect:Discuss} is the discussion
and outlook. Appendix \ref{sect:Convention} gives the conventions we
use in our paper.

\section{One-loop matrix model effective vertices and dilatation
operators\label{sect:Matrix-oneloop}}
Let us recall the relation between matrix model vertices and dilatation
operators in \cite{Beisert:2002ff}. The one-loop two-point function can be
written as 
\begin{equation}\label{eq:two-point}
\langle {\cal O}_\alpha(x)\bar{\cal O}_{\bar \beta}(0)\rangle=\frac
1{|x|^{2(J+2)}}\Big(S_{\alpha\bar\beta}
+T_{\alpha\bar\beta}\log(|x\Lambda|^{-2})\Big)\,.
\end{equation}
The dilatation operator matrix element can be expressed as 
\begin{equation}
D_\alpha{}^{\beta}=(J+2)\delta_{\alpha}{ }^{\beta}+T_{\alpha\bar
\gamma}(S^{-1})^{\bar\gamma\beta}\,.
\end{equation}
The basic idea of \cite{Beisert:2002bb} is to construct matrix model
effective vertices $H$ to reproduce the one-loop $T$ matrix
elements for scalar sector by
\begin{equation}
T_{\alpha\bar\beta}=\langle {\cal O}_\alpha~ H ~\bar{\cal
O}_{\bar\beta}\rangle=H_{\alpha}{ }^{\gamma}\langle {\cal
O}_\gamma{\cal O}_{\bar\beta}\rangle=H_{\alpha}{
}^{\gamma}S_{\gamma\bar\beta}
\end{equation}
and 
\begin{eqnarray}
H&=&\frac{g^2}{2(4\pi^2)}\left(
V_D+V_F+V_K\right)
\label{eq:Boson-V}
\end{eqnarray}
with 
\begin{eqnarray}
V_D&=&\frac 12:{\rm Tr}[\phi_m^+,\phi_m^-][\phi_n^+,\phi_n^-]:
+N:{\rm Tr}\phi_m^+\phi_m^-: 
- :{\rm Tr}(\phi_m^+){\rm Tr}(\phi_m^-):
\label{eq:VD}
\\V_F&=&-
\frac 12:{\rm Tr}[\phi_m^+,\phi_n^+][\phi_m^-,\phi_n^-]:
\\V_K&=&
-\frac 14:{\rm Tr}[\phi_m^+,\phi_n^-][\phi_m^+,\phi_n^-]:
\end{eqnarray}
where repeated $m$ and $n$ sum over $1,\dots, 6$.  $\phi^+$ and
$\phi^-$ denote the $\phi$ at $x$ and  $0$ respectively and
contractions are nonzero only between $\phi^+$ and $\phi^-$:
\begin{eqnarray}
\langle
(\phi_i^+)_{ab}(\phi_j^-)_{cd}\rangle=\delta_{ij}\delta_{ac}\delta_{bd}
\,,\quad
\langle\phi_i^+\phi_j^+\rangle=\langle\phi_i^+\phi_j^+\rangle =0\,.
\end{eqnarray}
It can be proven that $V_D$ is cancelled out when acting on traces of
scalars $\phi$. But we need to reexamine this result when considering
operators with fermion pairs like $\Psi^{p,J}_{ij}$ later.  From
(\ref{eq:two-point}), we see that $H_{\alpha}{ }^{\gamma}$ are the
matrix elements of one-loop Dilation operator when acting on the
operators with traces of scalars. For the operators $\Psi^{p,J}_{ij}$
with a scalar fermion pair the forgoing discussion does not need to be
modified except that the $H$ operator should include the fermionic
sector. Notice that for states in (\ref{eq:operators}), $S_{\alpha
\bar\beta}$ is diagonalized and normalized. So to change the operators
into Matrix model operators we need to make the replcement:
\begin{eqnarray}
2\pi\phi_i\rightarrow \phi_i\,.
\end{eqnarray}

Since we are interested in the operators with only scalar fermion
pairs, we need only to calculate the correlation functions with scalar
fermion pairs in order to obtain the effective vertices. The
$\sigma^\mu$ matrices in the fermion propagators of SUSY YM
would be assembled into matrix model coefficients and will not
appear in the correlation functions of two fermions in the matrix
model.  We can
specify the contraction rules of the fermions in the matrix model: 
\begin{eqnarray}
\langle (\psi_\alpha)_{ab}(\bar \psi_{\dot
\alpha})_{cd}\rangle=\epsilon_{\alpha\dot\alpha}\delta_{ad}\delta_{bc}
\end{eqnarray}
and require 
\begin{eqnarray}
\psi_\alpha^\dagger=\bar\psi^{\dot\alpha}\,,
\quad
(\psi^\alpha)^\dagger=-\bar\psi_{\dot\alpha}\,.  
\end{eqnarray}
The correspondence between the SUSY YM operators and matrix model
operators is:
\begin{equation}
\left(2^{1/4}\pi\right)\psi\to\left(\frac1 {2^{1/4}}\right)\psi \,.
\end{equation}

We need to calculate the one-loop diagrams in SUSY YM in figure
\ref{fig:one-loop}
\begin{figure}[h]
\begin{center}
\begin{picture}(20,40)(0,-20)
\DashLine(10,-20)(10,20){3}
\CArc(10,0)(8,-90,90)
\end{picture}
\hspace{1cm}
\begin{picture}(20,40)(0,-20)
\DashLine(10,-20)(10,20){3}
\GlueArc(10,0)(8,-90,90){2}{4}
\end{picture}
\hspace{1cm}
\begin{picture}(20,40)(0,-20)
\DashLine(0,-20)(0,20){3}
\Line(20,-20)(20,20)
\Gluon(0,0)(20,0){3}{3}
\end{picture}
\hspace{1cm}
\begin{picture}(20,40)(0,-20)
\DashLine(0,20)(0,0){3}
\DashLine(20,0)(0,-20){3}
\DashLine(0,0)(20,0){3}
\Line(20,-20)(0,0)
\Line(20,20)(20,0)
\end{picture}
\hspace{1cm}
\begin{picture}(20,40)(0,-20)
\DashLine(0,-20)(10,-10){3}
\DashLine(0,20)(10,10){3}
\DashLine(10,-10)(10,10){3}
\Line(20,-20)(10,-10)
\Line(20,20)(10,10)
\end{picture}
\hspace{1cm}
\begin{picture}(20,40)(0,-20)
\DashLine(0,-20)(0,20){3}
\DashLine(20,-20)(20,20){3}
\Gluon(0,0)(20,0){3}{3}
\end{picture}
\hspace{1cm}
\begin{picture}(20,40)(0,-20)
\DashLine(0,-20)(10,-10){3}
\DashLine(0,20)(10,10){3}
\Line(10,-10)(10,10)
\DashLine(20,-20)(10,-10){3}
\DashLine(20,20)(10,10){3}
\end{picture}\end{center}
\caption{One-loop diagrams in calculation of the effective vertices.
The scalars and fermions are denoted as solid and dashed lines
respectively.\label{fig:one-loop}}
\end{figure}
to get the effective vertices. One can consult \cite{Dobashi:2006fu}
for a detailed calculation of these diagrams.  Because we are
considering operators with only one scalar fermion pair, we do not
need to worry about the contractions between different scalar fermion
pairs. For a connected diagram like:
\begin{eqnarray}
\begin{array}{l}
\begin{picture}(20,40)(0,-20)
\DashLine(0,-20)(0,20){3}
\Line(20,-20)(20,20)
\Gluon(0,0)(20,0){3}{3}
\end{picture}\end{array}
\sim\frac{4ig^2}{(4\pi^2)^3}(\bar\sigma^\mu)^{\dot\beta\alpha}\frac{(x-y)_\mu\ln(-(x-y)^{-2}\Lambda^{-2})}{(x-y)^6}\,,
\end{eqnarray}
we can  add another free fermion propagator to combine with the
fermion $su(2)$ indices to form a Lorentz scalar coefficient. As a
result we include another trace of a fermion pair in the matrix model
vertices. The coefficients are obtained so that the $T$ matrix is the
same both from SUSY YM side and the matrix model side:
\begin{eqnarray}
H_{\psi}=\frac {g^2 }{2(4\pi^2)}\left\{N:\left({\rm
Tr}\psi^{\alpha-}\bar\psi^{\dot\beta+}-\frac 1 N{\rm
Tr}\psi^{\alpha-}{\rm Tr}\bar\psi^{\dot\beta+}\right){\rm
Tr}\psi_{\alpha}^{-}\bar\psi_{\dot\beta}^{+}:+(+\leftrightarrow -)
\right.\nonumber\\
-\frac 1 4:{\rm Tr}\left(\{\psi^{\beta A-},\bar \psi^{{\dot \beta}+}_A\}
[\bar Z^+,Z^-]\right) {\rm Tr}\psi_{\beta}^{-}\bar\psi_{\dot\beta}^{+}:+(+\leftrightarrow -)
\nonumber\\
-\frac 1 4:{\rm Tr}\left(\{\psi^{\beta A+},\bar \psi^{{\dot \beta}-}_A\}
[\bar Z^+,Z^-]\right) {\rm Tr}\psi_{\beta}^{+}\bar\psi_{\dot\beta}^{-}: +(+\leftrightarrow -) 
\nonumber\\
+\frac 1 2\sum_{i=1,2}:{\rm Tr}\left([\psi^{\beta i-},\bar Z^+]
[Z^-,\bar \psi^{{\dot \beta}+}_i]\right) {\rm
Tr}\psi_{\beta}^{B-}\bar\psi_{\dot\beta B}^{+}: +(+\leftrightarrow -)
\nonumber\\
-\frac 1 2\sum_{i=3,4}:{\rm Tr}\left([\psi^{\beta i+}, Z^-]
[\bar Z^+,\bar \psi^{{\dot \beta}-}_i]\right) {\rm
Tr}\psi_{\beta}^{B+}\bar\psi_{\dot\beta B}^{-}: +(+\leftrightarrow -) 
\nonumber\\
+\frac 1 4\sum_{i=1,2}:{\rm Tr}\left([\psi^{\beta i+},\bar Z^+]
[Z^-,\bar \psi^{{\dot \beta}-}_i]\right) {\rm
Tr}\psi_{\beta}^{B-}\bar\psi_{\dot\beta B}^{+}: +(+\leftrightarrow -)
\nonumber\\
-\frac 1 4\sum_{i=3,4}:{\rm Tr}\left([\psi^{\beta i-}, Z^-]
[\bar Z^+,\bar \psi^{{\dot \beta}+}_i]\right) {\rm
Tr}\psi_{\beta}^{B+}\bar\psi_{\dot\beta B}^{-}: +(+\leftrightarrow -)   
\nonumber\\
+\frac 1 2:{\rm Tr}\left(\{\psi^{\beta A+},\bar \psi^{{\dot \beta}-}_A\}
\{\psi_{\beta}^{B+},\bar \psi_{\dot \beta B}^{-}\}\right):  +(+\leftrightarrow -) 
\nonumber\\
\left.-\frac 1 4:{\rm Tr}\left(\{\psi^{\beta A+},\psi^{B+}_{\beta}\}
\{\bar\psi_{\dot\beta A}^{-},\bar \psi^{\dot \beta-}_{B}\}\right):  +(+\leftrightarrow -) 
\right\}\,,
\label{eq:Fermion-V}
\end{eqnarray}
where $(+\leftrightarrow -)$ means exchanging the $+$ and $-$ in the
superscripts. We will use this notation throughout the paper. 

Acting on an operator $\Psi_{ij}^{p,J}$, the terms in
(\ref{eq:Boson-V}) which contribute are only in (\ref{eq:VD}) and can
be written as:
\begin{eqnarray}\label{eq:VD-Z}
V_D\sim\frac 1 2 {\rm Tr}\left([\bar Z^+,Z^-][\bar
Z^+,Z^-]\right)-N{\rm Tr} Z^-\bar Z^++{\rm Tr} Z^-{\rm Tr}\bar Z^+
\end{eqnarray}
Like in the bosonic case, we can prove that these terms cancel with
the second and third line in (\ref{eq:Fermion-V}). To see this, contract
one $\bar Z^+$ in the first term of (\ref{eq:VD-Z}) with $Z$ in ${\rm Tr}(\psi^{i,\alpha} Z^p
\psi^j_{\alpha}Z^{J-p})$ and after some cancellations, the terms left are
\begin{eqnarray}
&&\frac12\left\{-{\rm Tr}(\psi^{i\alpha-}[\bar Z^+,Z^-]Z^p\psi^{j-}_\alpha
Z^{J-p})+\frac12{\rm Tr}(\psi^{i\alpha-}Z^p[\bar Z^+,Z^-]\psi^{j-}_\alpha Z^{J-p})
\right.\nonumber \\
&&\left.-{\rm Tr}(\psi^{i\alpha-}Z^p\psi^{j-}_\alpha[\bar Z^+,Z^-]
Z^{J-p})]
+\frac12{\rm Tr}(\psi^{i\alpha-}Z^p\psi^{j-}_\alpha
Z^{J-p}[\bar Z^+,Z^-])\right\}
\nonumber \\
&=&-\frac12{\rm Tr}([\psi^{i\alpha-},[\bar Z^+,Z^-] ]Z^p\psi^{j-}_\alpha
Z^{J-p})-\frac12{\rm Tr}(\psi^{i\alpha-}Z^p[\psi^{j-}_\alpha,[\bar Z^+,Z^-]]
Z^{J-p})\,.\label{eq:cancel}
\end{eqnarray}
Contracting the first term in the second line of (\ref{eq:Fermion-V})
with the same operator and keeping track of the minus sign when one
fermionic operator moves over the other, one gets   
\begin{eqnarray}
\frac12{\rm Tr}([\psi^{i\alpha-},[\bar Z^+,Z^-] ]Z^p\psi^{j-}_\alpha
Z^{J-p})+\frac12{\rm Tr}(\psi^{i\alpha-}Z^p[\psi^{j-}_\alpha,[\bar
Z^+,Z^-]]Z^{J-p})\,,
\end{eqnarray}
which just cancels (\ref{eq:cancel}). Also as in the bosonic case,
contracting the $\bar Z^+$ which is used in the contraction in
(\ref{eq:cancel}) with the $Z^-$ inside the first term in
(\ref{eq:VD-Z}) cancels the other terms left.

\section{One-loop operator mixing\label{sect:Mixing-oneloop}}
We only need to discuss the operator mixing at planar level. The 
pure bosonic operator eigenstates are given in \cite{Beisert:2002tn}.
Since we confine ourselves to only two impurity cases, it is easier to
directly diagonalize the $H$ matrix than use a Bethe ansatz. Let us
review the bosonic operator mixing first. ${\cal O}_{ij}$ can be
decomposed into $SO(4)$ representations:
\begin{eqnarray}
{\cal O}^{-1,J}_1&=&-{\rm Tr}(\bar Z Z^{J+1})\,,
\nonumber\\
{\cal O}^{p,J}_1&=&\frac 1 2 \sum_{i=1,2,4,5} {\cal O}^{p,J}_{ii}\,,
\nonumber\\
{\cal O}^{p,J}_{(ij)}&=&\frac12\left({\cal O}^{p,J}_{ij}+{\cal
O}^{p,J}_{ji}\right)-\frac 1 2 {\cal
O}^{p,J}_{1}\,,
\nonumber\\
{\cal O}^{p,J}_{[ij]}&=&\frac12\left({\cal O}^{p,J}_{ij}-{\cal
O}^{p,J}_{ji}\right)\,.
\end{eqnarray}
${\cal O}^{-1,J}_{1}$ and ${\cal O}^{p,J}_{1}$ are $SO(4)$ singlets.
${\cal O}^{p,J}_{(ij)}$ is in the $9$ representation and ${\cal
O}^{p,J}_{[ij]}$ is in the $3+\bar 3$ representation. 

The $H$ matrix for
${\cal O}^{p,J}_1, (p=-1,0,1,\dots,[J/2])$ is 
\begin{eqnarray}\label{eq:H-O-single}
\begin{array}{ll}
\text{for } J \text { odd:}&\text{for } J \text{ even:}
\cr
\frac{g^2N}{4\pi^2}\left(\begin{array}{ccccccc} 
1&-1&&&&&
\cr
-1&2&-1&&&&\cr &-1&2&-1&&&\cr&&&\cdots&\cdots&&\cr&&&&-1&2&-1\cr&&&&&-1&1
\end{array}
\right)\,,\hspace{1.5cm}&
\frac{g^2N}{4\pi^2}
\left(\begin{array}{ccccccc} 
1&-1&&&&&
\cr
-1&2&-1&&&&\cr &-1&2&-1&&&\cr&&&\cdots&\cdots&&\cr&&&&-1&2&-2\cr&&&&&-1&2
\end{array}\right)\,,
\cr
\text{Eigenstates: } n=0,\cdots, (J+1)/2 &\text{Eigenstates: }n=0,\cdots, J/2+1
\cr
\begin{array}{ll}
\left( 2\cos\frac{\pi n}{J+3},\dots,\right.& 2\cos\frac{(2p+3)\pi
n}{J+3},\cr&\left.\dots,2\cos\frac{(J+2)\pi
n}{J+3}\right)^T \,,
\end{array}
&
\begin{array}{ll}
\left( 2\cos\frac{\pi n}{J+3}, \dots\right.,&2\cos\frac{(2p+3)\pi
n}{J+3},\dots,
\cr&\left.2\cos\frac{(J+1)\pi
n}{J+3},\cos \pi n\right)^T\,,
\end{array}\end{array}
\end{eqnarray}
where all the unspecified matrix elements are zero. The eigenstates
can be written in both cases as
\begin{eqnarray}
\label{eq:Eigen-O-single}
\tilde {\cal O}_1^{J,n}=\sum_{p=0}^{J}\cos\frac{(2p+3)\pi
n}{J+3}\ {\cal O}_1^{p,J}+2\cos \frac {\pi n}{J+3}\ {\cal
O}^{-1,J}_1\,.
\end{eqnarray}
The eigenvalue is 
\begin{equation}
\delta D_2=\frac{g^2N}{4\pi^2}4\sin^2\frac{ n\pi}{J+3}\,.
\end{equation}
In fact, we could take the coefficients before ${\cal O}_1^{p,J}$ to
be $\cos\frac{(2p+a)n\pi}{J+b}$ or $\sin\frac{(2p+a)n\pi}{J+b}$ which
satisfies the eigenvalue requirement except the last line and the
first line in the matrix for $J=$ odd. The last line and the first
line act as boundary conditions: the last line determines the
coefficients to be $\cos\frac{(2p+a)n\pi}{J+a}$ and if we require $a$, $b$
to be independent of $J$, the first line determines $a=3$.

For the symmetric operator ${\cal O}_{(ij)}$, the $H$ matrix is the
same form as in the ${\cal O}_1$ case except that $p$ starts from $0$ not from
$-1$. As a result, the eigenstates are:
\begin{eqnarray}\label{eq:Eigen-O-sym}
\tilde {\cal O}_{(ij)}^{J,n}=\sum_{p=0}^{J}\cos\frac{(2p+1)\pi
n}{J+1}\ {\cal O}_{(ij)}^{p,J}\,,
\end{eqnarray}
and the eigenvalue is 
\begin{equation}
\delta D_2=\frac{g^2N}{4\pi^2}4\sin^2\frac{ n\pi}{J+1}.
\end{equation}

For the anti-symmetric operators ${\cal O}_{[ij]}^{p,J}$,
$(p=0,1,\dots,[(J-1)/2])$, the $H$ matrix is
\begin{eqnarray}\label{eq:H-O-antisym}
\begin{array}{ll}
\text{for } J \text { odd:}&\text{for } J \text{ even:}
\cr
\frac{g^2N}{4\pi^2}\left(\begin{array}{ccccccc} 
2&-1&&&&&
\cr
-1&2&-1&&&&\cr &-1&2&-1&&&\cr&&&\cdots&\cdots&&\cr&&&&-1&2&-1\cr&&&&&-1&3
\end{array}
\right)\,,\hspace{1.5cm}&
\frac{g^2N}{4\pi^2}
\left(\begin{array}{ccccccc} 
2&-1&&&&&
\cr
-1&2&-1&&&&\cr &-1&2&-1&&&\cr&&&\cdots&\cdots&&\cr&&&&-1&2&-1\cr&&&&&-1&2
\end{array}\right)\,,
\cr
\text{Eigenstates: } n=1,\cdots, (J+1)/2 &\text{Eigenstates:
}n=1,\cdots, J/2\,,
\cr
\begin{array}{ll}
\left( 2\sin\frac{2\pi n}{J+2},\dots,\right.& 2\sin\frac{(2p+2)\pi
n}{J+2},\cr&\left.\dots,2\sin\frac{(J+1)\pi
n}{J+2}\right)^T \,,
\end{array}
&
\begin{array}{ll}
\left( 2\sin\frac{2\pi n}{J+2}, \dots\right.,&2\sin\frac{(2p+2)\pi
n}{J+2},
\cr&\left.\dots,2\sin\frac{J\pi
n}{J+2}\right)^T\,,
\end{array}\end{array}
\end{eqnarray}
The eigenstates can be written as 
\begin{eqnarray}\label{eq:Eigen-O-antisym}
\tilde {\cal O}_{[ij]}^{J,n}=\sum_{p=0}^{J}\sin\frac{(2p+2)\pi
n}{J+2}\ {\cal O}_{[ij]}^{p,J}
\end{eqnarray}
and the eigenvalue is 
\begin{equation}
\delta D_2=\frac{g^2N}{4\pi^2}4\sin^2\frac{ n\pi}{J+2}\,.
\end{equation}

For operators with a scalar fermion pair, it can be checked that the
Hamiltonian can be recast  as
\begin{eqnarray}
H=-\frac {g^2N }{2(4\pi^2)}\sum_{i=1}^{L}(1-\Pi_{i,i+1})
\end{eqnarray} 
where $\Pi_{i,i+1}$ is the graded permutation operator which exchanges
the adjacent fields and picks up a minus sign if the two fields are
fermions. $L$ is the total number of fields in the operator to be
acted on. This equation may not be true for two or more pairs of
fermions, because there  could be changes like $\psi^\alpha
\psi_\alpha\psi^\beta \psi_\beta\to\psi^\alpha \psi^\beta
\psi_\beta\psi_\alpha $. 

As in the pure bosonic cases, we can also define symmetric and
antisymmetric combinations for $(i,j= 1,2)$ or $(i,j=3,4)$:
\begin{eqnarray}
\Psi_{(ij)}^{p,J}=\frac 1 2 (\Psi_{ij}^{p,J}+\Psi_{ji}^{p,J})\,,
\nonumber \\
\Psi_{[ij]}^{p,J}=\frac 1 2 (\Psi_{ij}^{p,J}-\Psi_{ji}^{p,J})\,.
\end{eqnarray}
$\Psi_{(ij)}^{p,J}$ are in the $3$ representation for $i,j=1,2$ or
$\bar 3$ representation for $i,j=3,4$. We expect that ${\cal
O}^{p,J}_{[ij]}$ will be mixed with $\Psi^{p,J-1}_{(ij)}$ and ${\cal
O}^{p,J}_{1}$ with $\Psi^{p,J-1}_{[ij]}$.  

The $H$ matrix for $\Psi^{p,J}_{(ij)}$ is the same as in ${\cal
O}^{p,J}_{(ij)}$ case, therefore the eigenstates could be obtained
just by replacing $\cal O$ with $\Psi$ in (\ref{eq:Eigen-O-sym}) and the
eigenvalues are the same. In a similar way, the $\Psi^{p,J}_{[ij]}$
case  is the same
as ${\cal O}^{p,J}_{[ij]}$ case. 

From these diagonalization procedures, we see that for a
matrix with $-1, 2,-1$ at the near diagonal and diagonal positions
like in (\ref{eq:H-O-single}) and (\ref{eq:H-O-antisym}), we can use
an ansatz like $\cos \frac{(2p+a)n\pi}{J+b}$ or $\sin
\frac{(2p+a)n\pi}{J+b}$ as coefficients because they automatically
satisfy the eigenvalue equations. Combining the coefficient before the
matrix (\ref{eq:H-O-single}) and (\ref{eq:H-O-antisym}), the
eigenvalue is $\frac{g^2N}{4\pi^2}4\sin^2\frac{ n\pi}{J+b}$. The
variables $a$, $b$ and $\sin$ or $\cos$ are determined from the
boundary condition: the first line and the last line of the matrix and
the requirement that $a$ and $b$ are independent of $J$. This is much
like solving an eigenvalue problem of a differential equation with
boundary conditions. This insight is helpful when we solve the
mixing between pure scalar operators and operators with fermions.

\section{Order analysis\label{sect:Order}}
In general, the $SO(4)$ singlet part of ${\cal O}^{p,J}_{1}$ and
$\Psi^{p,J-1}_{[ij]}$
should also be mixed with operator
\begin{eqnarray}
V^{p,J}={\rm Tr}(D_\mu Z Z^p D^\mu Z Z^{J-p-2})\,.
\end{eqnarray}
In paper \cite{Georgiou:2009tp}, the authors use the method developed
in \cite{Georgiou:2008vk} to discuss the mixings among these operators
in which $V^{p,J}$ appears only from $O(g^2)$ terms. We will try to
understand the general form of the eigenstate vector by order analysis
in this section.

Let us look at the Dilatation matrix structure of these three kinds of
operator system to $O(g^4)$.  For the combined operator basis
$(\dots,{\cal O}^{p,J},\dots;\dots \Psi^{p,J-1},\dots; \dots,
V^{p,J-2},\dots)$, the correction to the dilatation operator up to
$O(g^4)$ in matrix blocks is in this form:
\begin{eqnarray}
\delta D\sim \left(\begin{array}{ccc}
O(g^2)+O(g^4) & O(g^3)        & O(g^4)\\
O(g^3)        & O(g^2)+O(g^4) & O(g^3)\\
O(g^4)        & O(g^3)        & O(g^2)+O(g^4)\\
\end{array}\right)\,.
\end{eqnarray}
In general, the eigenvector should look like this
\begin{eqnarray}
\sum A_p{\cal O}^{p,J}+\sum B_p \Psi^{p,J-1} + \sum C_p V^{p,J-2}\,.
\end{eqnarray}
According to general perturbation theory of quantum mechanics, we first
diagonalize the $O(g^2)$ part of the matrix. This is done in previous
section and in paper \cite{Beisert:2002tn}. The three groups of
eigenvectors correspond to $A_p\neq 0$, $B_p=C_p=0$ or $B_p\neq 0$,
$A_p=C_p=0$ or $C_p\neq 0$, $A_p=B_p=0$ and the eigenvalues are not
degenerate. We will use $\tilde {\cal O}^{J,n}$, $\tilde \Psi^{J-1,n}$
and $\tilde V^{J-2,n}$ to denote these eigenstates. So the
non-degenerate perturbation theory could be used to calculate the
correction to the next order $O(g^3)$. It is easy to see that there is
no correction to the eigenvalues to this order since the correction
terms are proportional to $\langle \tilde {\cal
O}^{J,n}|H(O(g^3))|\tilde {\cal O}^{J,n}\rangle$, $\langle \tilde
\Psi^{J,n}|H(O(g^3))|\tilde \Psi^{J,n}\rangle$ or $\langle \tilde
V^{J,n}|H(O(g^3))|\tilde V^{J,n}\rangle$. If we are only interested in
the mixing between ${\cal O}^{p,J}$ and $\Psi^{p,J-1}$, we can just
take the upper left $2\times2$ matrix and diagonalize it. This is what
we will do later.  We will see that it is easier to do it without
using the quantum mechanics perturbation theory result.  We can then
do the perturbation to $O(g^4)$. Just like the $O(g^3)$ case, the
nondiagonal matrix blocks do not contribute to the eigenvalue but
contribute to the mixing coefficients. This means that if we only
consider the lowest order of the mixing coefficients between ${\cal
O}^{p,J}$ and $V^{p,J-2}$ we can just use the diagonal $O(g^2)$ blocks
and nondiagonal $O(g^4)$ blocks.  Combining $O(g^3)$ and $O(g^4)$
results, the three kinds of solutions for the coefficients should be
$A_p\sim O(1)$, $B_p\sim O(g)$, $C_p\sim O(g^2)$ or $A_p\sim O(g)$,
$B_p\sim O(1)$, $C_p\sim O(g)$ or $C_p\sim O(g^2)$, $A_p\sim O(g)$,
$B_p\sim O(1)$. Notice that the finite $J$ condition plays an
important role here, because in the infinite $J$ limit the three
groups of the leading order eigenstates degenerate and the
non-degenerate perturbation argument can not be used here. Just
diagonalizing the upper left $2\times 2$ block matrix would  not be
valid.

\section{Effective vertices of $O(g^3)$\label{sect:Matrix-O3}} 
In order to study the mixing between ${\cal O}^{p,J}_{ij}$ and
$\Psi^{p,J-1}_{ij}$ we need to calculate dilatation operator changing
fermions into bosons and vice versa.  The lowest order nonzero diagrams
are of $O(g^3)$:
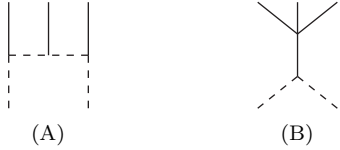
\begin{center}
\begin{picture}(30,55)(0,-35)
\DashLine(0,-20)(0,0){3}
\DashLine(0,0)(30,0){3}
\DashLine(30,0)(30,-20){3}
\Line(0,0)(0,20)
\Line(15,0)(15,20)
\Line(30,0)(30,20)
\Text(15,-30)[c]{\scriptsize(A)}
\end{picture}
\hspace{2cm}
\begin{picture}(30,55)(0,-35)
\DashLine(0,-20)(15,-8){3}
\DashLine(15,-8)(30,-20){3}
\Line(15,8)(15,-8)
\Line(15,8)(0,20)
\Line(15,8)(15,20)
\Line(15,8)(30,20)
\Text(15,-30)[c]{\scriptsize(B)}
\end{picture}
\end{center}
It is easy to calculate the divergent parts of these two diagrams using
relation (B.1) in \cite{Georgiou:2008vk} and differential
regularization \cite{Freedman:1991tk}. The effective vertices can be
obtained in the same spirit as in section \ref{sect:Matrix-oneloop}: 
\begin{eqnarray}
H_{(A)}&=&-\frac{g^3}{(2\pi)^3}{\rm Tr}\Big[\big\{ [\psi^{\alpha
A-},\Phi_{AB}^{+}],[\Phi_{CD}^+,\psi_\alpha^{D-}]\big\}\Phi^{BC+}-\big\{[\bar\psi_{\dot\alpha
A}^-,\Phi^{AB+}],[\Phi^{CD+},\bar\psi^{\dot\alpha-}_D]\big\}\Phi_{BC}^+\Big]
\nonumber
\\
&&+(+\leftrightarrow -)
\label{eq:H-O3-A}
\\
H_{(B)}&=&-\frac{g^3}{8(2\pi)^3}{\rm Tr}\Big[-\{ \psi^{\alpha
A-},\psi_\alpha^{B-}\}\big[\Phi^{CD+},[\Phi^{AB+},\Phi_{CD}^-]\big]+\{\bar
\psi^-_{\dot\alpha
A},\bar\psi^{\dot\alpha-}_B\}\big[\Phi_{CD}^+,[\Phi^{AB+},\Phi^{CD+}]\big]\Big]
\nonumber
\\
&&+(+\leftrightarrow -)
\label{eq:H-O3-B}
\end{eqnarray}
Using (\ref{eq:Z-Phi}), we can change  $\Phi$ into $Z$ and $\bar Z$:
\begin{eqnarray}
H_{(A)}&=&-\frac{g^3}{(2\pi)^3}{\rm Tr}\Big[-\frac 1
8\epsilon_{ijk}[\psi^{\alpha i-},\bar Z^{j+}]\big(-\big[ [\bar
Z^{l+},\psi^{k-}_\alpha],Z^+_l\big]-\big[ [
Z^+_l,\bar Z^{k+}],\psi^{l-}_\alpha\big] +\big[ [
\psi^{l-}_\alpha,Z^+_l],\bar Z^{k+}\big]\big)
\nonumber \\
&&+\frac 1 4 [\psi^{\alpha 4-},Z^+_j]\big(\big[ [\bar
Z^{i+},\psi^{j-}_\alpha],Z^+_i\big]-\big[ [
\bar Z^{i+},Z^+_i],\psi^{i-}_\alpha\big] 
\big)
-\frac 1 8\epsilon^{ijk}\big\{[\psi^{\alpha
4-},Z^+_i],[Z^+_j,\psi^{4-}_{\alpha}]\big\}Z^+_k
\nonumber\\
&&+h.c.\Big]+(+\leftrightarrow -)
\label{eq:H-O3-Z-A}
\\
H_{(B)}&=&-\frac{g^3}{8(2\pi)^3}{\rm Tr}\left[-\left(\frac 1
4\epsilon_{ijk}\{\psi^{\alpha i-},\psi^{j-}_\alpha\}+\frac 12\{\bar
\psi^-_{\dot\alpha 4},\bar
\psi^{\dot\alpha-}_k\}\right)\left(\big[\bar Z^{l+},[\bar
Z^{k+},Z^+_l]\big]+\big[ [\bar Z^{l+},\bar Z^{k+}],Z^+_l\big]\right)
\right.\nonumber \\
&&+h.c.\Big]+(+\leftrightarrow -)
\label{eq:H-O3-Z-B}
\end{eqnarray} 
Note that the last term of the second line in
(\ref{eq:H-O3-Z-A}) is just the $H_3$ part for the subsector $su(2|3)$
discussed in \cite{Beisert:2003ys}. 

We only consider planar diagrams in this paper, so the
terms we are interested in must have adjacent fermions and bosons, and
only $\Psi^{p=0,J}_{ij}$ has non-zero correlation function with
${\cal O}^{p=0}_{ij}$ or ${\cal O}^{p=1}_{ij}$ at planar level to $O(g^3)$. The
matrix model vertices must involve all the  impurities in both
operators in the two-point correlation function.  So, using the cyclic
property of the trace of operator string, for effective vertices like
${\rm Tr}(\psi_l^-\psi_m^-\phi_i^+\phi_j^+ \bar Z^+)$ in which the two
impurity bosonic fields are adjacent, we can change the position of
$Z$ or $\bar Z$ from the end of the three bosonic fields to the
beginning of them or vice versa without changing the two point correlation 
function, that is ${\rm Tr}(\psi_l^-\psi_m^-\phi_i^+\phi_j^+ \bar
Z^+)\leftrightarrow {\rm Tr}(\psi_l^-\psi_m^-\bar Z^+\phi_i^+\phi_j^+
)$. Note that this operation only works for two impurity operators.  By
appropriately using this operation, we collect the terms we are
interested in as:
\begin{eqnarray}
H_{(A)}+H_{(B)}&\sim& \frac{g^3}{4(2\pi)^3}{\rm Tr}\Bigg\{-\bigg(\sum_{l=1,2,4,5}
\phi^-_l\phi^-_lZ^-+2Z^-Z^-\bar Z^-\bigg)
(\bar\psi^+_{[1\dot
\alpha}\bar\psi_{2]}^{\dot \alpha+}-\psi^{[3
\alpha+}\psi^{4]+}_{\alpha}) 
\nonumber \\
&&+i\Big(2\phi^-_{[1}\phi^-_{4]}Z^--2\phi^-_{[2}\phi^-_{5]}Z^--\phi^-_{[1}Z^-\phi^-_{4]}+\phi^-_{[2}Z^-\phi^-_{5]}\Big)
(\bar\psi^+_{1\dot
\alpha}\bar\psi_{2}^{\dot \alpha+}+\bar\psi^+_{2\dot
\alpha}\bar\psi_{1}^{\dot \alpha+}) 
\nonumber \\
&&-\Big(2\phi^-_{[1}\phi^-_{2]}Z^-+2\phi^-_{[4}\phi^-_{5]}Z^--\phi^-_{[1}Z^-\phi^-_{2]}-\phi^-_{[4}Z^-\phi^-_{5]}\Big)
(\bar\psi^+_{1\dot
\alpha}\bar\psi_{1}^{\dot \alpha+}+\bar\psi^+_{2\dot
\alpha}\bar\psi_{2}^{\dot \alpha+}) 
\nonumber \\
&&-i\Big(2\phi^-_{[1}\phi^-_{5]}Z^-+2\phi^-_{[2}\phi^-_{4]}Z^--\phi^-_{[1}Z^-\phi^-_{5]}-\phi^-_{[2}Z^-\phi^-_{4]}\Big)
(\bar\psi^+_{1\dot
\alpha}\bar\psi_{1}^{\dot \alpha+}-\bar\psi^+_{2\dot
\alpha}\bar\psi_{2}^{\dot \alpha+})
\nonumber \\
&&-i\Big(2\phi^-_{[1}\phi^-_{4]}Z^-+2\phi^-_{[2}\phi^-_{5]}Z^--\phi^-_{[1}Z^-\phi^-_{4]}-\phi^-_{[2}Z^-\phi^-_{5]}\Big)
(\psi^{3
\alpha+}\psi^{4+}_{\alpha}+\psi^{4
\alpha+}\psi^{3+}_{\alpha}) 
\nonumber \\
&&+\Big(2\phi^-_{[1}\phi^-_{2]}Z^--2\phi^-_{[4}\phi^-_{5]}Z^--\phi^-_{[1}Z^-\phi^-_{2]}+\phi^-_{[4}Z^-\phi^-_{5]}\Big)
(\psi^{3
\alpha+}\psi^{3+}_{\alpha}+\psi^{4
\alpha+}\psi^{4+}_{\alpha}) \nonumber \\
&&-i\Big(2\phi^-_{[1}\phi^-_{5]}Z^--2\phi^-_{[2}\phi^-_{4]}Z^--\phi^-_{[1}Z^-\phi^-_{5]}+\phi^-_{[2}Z^-\phi^-_{4]}\Big)
(\psi^{3
\alpha+}\psi^{3+}_{\alpha}-\psi^{4
\alpha+}\psi^{4+}_{\alpha})
 \nonumber\\
&&+(h.c., +\leftrightarrow -)\Bigg\}\,,
\label{eq:D-O3}
\end{eqnarray}
where $\phi_{[i}\phi_{j]}=\frac 1 2(\phi_i\phi_j-\phi_j\phi_i)$. We can define:
\begin{xalignat}{2}
&{\cal O}^{p,J}_{[ij,kl]}={\cal O}^{p,J}_{[ij]}-{\cal
O}^{p,J}_{[kl]}\,,
&&{\cal O}^{p,J}_{(ij,kl)}={\cal O}^{p,J}_{[ij]}+{\cal
O}^{p,J}_{[kl]}\,,
\nonumber \\
&\Psi^{p,J}_{(ij,kl)}=\frac12(\Psi^{p,J}_{ij}+\Psi^{p,J}_{kl})\,,
&&\Psi^{p,J}_{[ij,kl]}=\frac12(\Psi^{p,J}_{ij}-\Psi^{p,J}_{kl})\,.
\end{xalignat}
From (\ref{eq:D-O3}) we see that ${\cal O}_1^{p,J}$ are mixed with
$\Psi_{[12]}^{p,J-1}$ and $\Psi_{[34]}^{p,J-1}$, ${\cal
O}_{[25,14]}^{p,J}$ mixed with $\Psi_{(12)}^{p,J-1}$, ${\cal
O}_{(12,45)}^{p,J}$ mixed with $\Psi_{(11,22)}^{p,J-1}$, etc. We can
list these mixed operators in the rows of  table \ref{tab:mixing}.

\begin{table}[h]
\begin{center}
\begin{tabular}{|c|c|}
\hline
${\cal O}_1^{p,J}$&$\Psi_{[12]}^{p,J-1}-\Psi_{[34]}^{p,J-1}$
\cr 
\hline
$i{\cal
O}_{[14,25]}^{p,J}$ & $\Psi_{(12)}^{p,J-1}$
\cr 
\hline
$-{\cal
O}_{(12,45)}^{p,J}$ & $\Psi_{(11,22)}^{p,J-1}$
\cr 
\hline
 $-i{\cal
O}_{(24,15)}^{p,J}$ & $\Psi_{[11,22]}^{p,J-1}$
\cr 
\hline
$-i{\cal
O}_{(14,25)}^{p,J}$ & $\Psi_{(34)}^{p,J-1}$
\cr 
\hline
${\cal
O}_{[12,45]}^{p,J}$ & $\Psi_{(33,44)}^{p,J-1}$
\cr 
\hline
 $-i{\cal
O}_{[15,24]}^{p,J}$ & $\Psi_{[33,44]}^{p,J-1}$
\cr
\hline
\end{tabular}
\end{center}
\caption{\label{tab:mixing}The operators on the same row will be mixed.}
\end{table}
These mixing patterns are consistent with group analysis as follows:
if we keep $Z=2\Phi_{34}$ and $\bar Z=2 \Phi_{12}$
invariant, the $SU(4)_R$ symmetry is broken to
$SO(4)=SU(2)\times SU(2)$ with $\psi^{1,2}$ and $\bar \psi_{3,4}$ as
fundamental representations of the two $SU(2)$'s respectively. Using (\ref{eq:Z-Phi}),
(\ref{eq:ZtoPhi}) and omitting $Z$, we can obtain:
\begin{eqnarray}
i{\cal O}_{[14,25]}\sim
\Phi^{14}\Phi^{23}-\Phi^{23}\Phi^{14}+\Phi^{24}\Phi^{13}-\Phi^{13}\Phi^{24}\,,
\\
-{\cal O}_{(12,45)}\sim
\Phi^{14}\Phi^{13}-\Phi^{23}\Phi^{24}-\Phi^{13}\Phi^{14}+\Phi^{24}\Phi^{23}\,,
\\
-i{\cal O}_{(24,15)}\sim
\Phi^{23}\Phi^{24}+\Phi^{14}\Phi^{13}-\Phi^{13}\Phi^{14}-\Phi^{24}\Phi^{23}\,,
\end{eqnarray}
from which we see that the indices $3$ and $4$ are antisymmetric and
form a singlet under the second $SU(2)$. The indices $1$ and $2$ have
precisely the same transformation property as the corresponding mixed
operators in table \ref{tab:mixing}. The same analysis can be
performed on the other rows. In fact, $\Psi_{11}$, $\Psi_{22}$ and
$\Psi_{(12)}$ form a $3$ representation of $SO(4)$ and $\Psi_{33}$,
$\Psi_{44}$ and $\Psi_{(34)}$ form a $\bar 3$ representation. The
corresponding scalars are also expected to form  $3$ and $\bar 3$
representions.  As a result, the operators $V^{p,J}$ are only mixed
with the first line of table \ref{tab:mixing} but not with the other
lines.

\section{Mixing of operators\label{sect:Mixing-O3}}

Let us look at the mixing between ${\cal O}_1$, $\Psi_{[12]}$ and
$\Psi_{[34]}$ first. Since there is no correlation between
$\Psi_{[12]}$ and $\Psi_{[34]}$ up to $O(g^3)$, we can consider the
mixing between ${\cal O}_1$ and $\Psi_{[12]}$, and  ${\cal O}_1$ and
$\Psi_{[34]}$ separately. Since the coefficients before $\bar
\psi_{[1}\bar\psi_{2]}$ and $\psi^{[3}\psi^{4]}$ in the vertices
differ only by a minus sign, the mixing coefficients for $\Psi_{[12]}$
and $\Psi_{[34]}$ with ${\cal O}_1$ must differ only by a minus sign.
For operator basis $({\cal O}^{-1,J}_1,{\cal O}^{0,J}_1,\dots, {\cal
O}^{[J/2],J}_1,\Psi^{0,J-1}_{[12]},\dots,\Psi^{[J/2]-1,J-1}_{[12]})$,
the $H$ matrix should combine the one-loop $H$ matrix for ${\cal
O}^{p,J}_1$
and $\Psi^{p,J-1}_{[12]}$ and some $O(g^3)$ crossing terms:
\begin{eqnarray}\label{eq:H-single12}
&&\hspace{-1cm}\text{for } J \text { odd:}
\nonumber\\
&&\frac{g^2N}{4\pi^2}\left(\begin{array}{cccccccccccc} 
1&-1&&&&&-\frac{g\sqrt{N}}{2\sqrt2(2\pi)}&&&&& \cr
-1&2&-1&&&&\frac{g\sqrt{N}}{2\sqrt2(2\pi)}&&&&& \cr
&&\cdots&\cdots&&&&&&&& \cr
&&&-1&2&-1&&&&&&\cr
&&&&-1&1&&&&&&\cr
-\frac{g\sqrt{N}}{\sqrt2(2\pi)}&\frac{g\sqrt{N}}{\sqrt2(2\pi)}&&&&&2&-1&&&&\cr
&&&&&&-1&2&-1&&&\cr
&&&&&&&&\cdots&\cdots&&\cr
&&&&&&&&&-1&2&-1\cr
&&&&&&&&&&-1&2
\end{array}
\right)
\label{eq:H-1-12}
\end{eqnarray}
After diagonalization, we require the non-diagonal elements to be
$O(g^4)$. So we require the correction to the eigenvector $\tilde{\cal
O}_1$ to be $O(g)$. The upper right non-diagonal element contribution
to the eigenvalue is already $O(g^4)$. The upper left block of the $H$
matrix acting on the $O(1)$ part gives just the one-loop case which determines
the eigenvalue to be $(g^2N/4\pi^2)4\sin^2(n\pi/(J+3))$. As stated at
the end of section \ref{sect:Mixing-oneloop}, we could make the
ansatz for the coefficients before $\Psi_{[12]}^{p,J-1}$ to be
$2gb\sin\frac{(2p+a)n\pi}{J+3}$. The boundary condition from the last
row of the $H$ matrix then determines $a=4$. The eigenvalue equation for
the coefficient before $\Psi^{0,J-1}$ acts as another boundary
condition which determines $b$. We then obtain the eigenvector with
$O(g)$ correction to $\tilde{\cal O}^{p,J}_1$: 
\begin{align}
Q_{1,12}^{J,n}=\tilde{\cal O}_1^{J,n}+\frac {\sqrt2g\sqrt N}{2\pi}\sin \frac
{n\pi}{J+3}\sum_{p=0}^{J-1}\sin\frac{(2p+4)n\pi}{J+3}&\,\Psi
^{p,J-1}_{[12]}\,,\nonumber\\
& \text{for } n=0,\dots,[J/2]+1\,.
\label{eq:Q1-12}
\end{align}

We can also require the coefficients before $\Psi_{[12]}^{p,J-1}$ to
be $O(1)$ and those before ${\cal O}_1^{J}$ to be $O(g)$. This time the
lower left nondiagonal element contribution is already $O(g^4)$, and
the upper right nondiagonal elements contribute to two eigenvalue
equations. As a result, the coefficient before
${\cal O}_1^{-1,J}$ should be considered as independent of the other
coefficients: we set it to be $A_{-1}$ and
for $p=0,\dots,(J-1)/2$ the coefficients before ${\cal O}^{p,J}_1$ are $2b
\cos \frac {(2p+a)n\pi}{J+1}$. As in previous case, the last line in the
upper left block determines $a=1$. Those two equations from the first
two lines determine $A_{-1}$ to be zero and
the value of $b$, thus the eigenvector in this case is 
\begin{align}
\Xi^{J-1,n}_{[12]}=\tilde\Psi_{[12]}^{J-1,n}-\frac{g\sqrt
N}{\sqrt2(2\pi)}\sin\frac
{n\pi}{J+1}\sum_{p=0}^{J}&\cos\frac{(2p+1)n\pi}{J+1}\,{\cal
O}_1^{p,J}\,,
\nonumber\\& \text{for } n=1,\dots,[(J+1)/2]\,.
\label{eq:Xi-12}
\end{align}
It is easy to check that for the $J=even$ case the eigenvectors can also
be written in this form.
Notice that the total number of eigenvectors in (\ref{eq:Q1-12}) and
(\ref{eq:Xi-12}) is equal to the rank of the matrix and the
eigenvalues are not degenerate. So these are all the eigenvectors of
the $H$ matrix (\ref{eq:H-single12}). 

For the mixing with $\Psi_{[34]}$, only the nondiagonal blocks in
(\ref{eq:H-1-12}) change signs. As a result the coefficients of $O(g)$
of the eigenvector change signs. So combined with (\ref{eq:Q1-12}), we
obtain the $O(g)$ correction to the eigenvector ${\cal O}_1^{p,J}$
\begin{eqnarray}
Q_{1}^{J,n}&=&\tilde{\cal O}_1^{J,n}+\frac {\sqrt2g\sqrt N}{2\pi}\sin \frac
{n\pi}{J+3}\sum_{p=0}^{J-1}\sin\frac{(2p+4)n\pi}{J+3}\left(\Psi
^{p,J-1}_{[12]}-\Psi^{p,J-1}_{[34]}\right)
\label{eq:Q1}
\end{eqnarray}
This is consistent with equation (3.17) in \cite{Georgiou:2008vk}.
(\ref{eq:Xi-12}) is consistent with the first two terms in equation
(2.12) in \cite{Georgiou:2009tp}. To obtain the full $O(g)$
correction, one must also consider the mixing matrix elements between
$\Psi_{[12]}$ and $D_\mu Z Z^p D^\mu ZZ^{J-p-1}$. 

As for the mixing between ${\cal O}^{p,J}_{[14,25]}$ and
$\Psi_{(12)}^{p,J-1}$, we can also have the $H$ matrix for operators
basis $\left(i{\cal O}^{0,J}_{[14,25]},\dots,i{\cal
O}^{[(J-1)/2],J}_{[14,25]},\Psi_{(12)}^{0,J-1},\dots,\Psi_{(12)}^{[(J-1)/2],J-1}\right)$:
\begin{eqnarray}\label{eq:H-[14,25]-(12)}
&&\hspace{-1cm}\text{for } J \text { odd:}
\\
&&\frac{g^2N}{4\pi^2}\left(\begin{array}{cccccccccccc} 
2&-1&&&&&\frac{g\sqrt{N}}{\sqrt2(2\pi)}&&&&& \cr
-1&2&-1&&&&-\frac{g\sqrt{N}}{2\sqrt2(2\pi)}&&&&& \cr
&&\cdots&\cdots&&&&&&&& \cr
&&&-1&2&-1&&&&&&\cr
&&&&-1&3&&&&&&\cr
\frac{\sqrt2 g\sqrt{N}}{(2\pi)}&-\frac{g\sqrt{N}}{\sqrt2(2\pi)}&&&&&1&-1&&&&\cr
&&&&&&-1&2&-1&&&\cr
&&&&&&&&\cdots&\cdots&&\cr
&&&&&&&&&-1&2&-2\cr
&&&&&&&&&&-1&2
\end{array}
\right)
\end{eqnarray}
It is easy to obtain the eigenvectors as before:
\begin{eqnarray}
Q^{J,n}_{[14,25]}&=&i\sum_{p=0}^J \sin \frac{(2p+2)n\pi}{J+2}{\cal
O}_{[14,25]}^{p,J}-\frac{g\sqrt N}{\sqrt 2
(2\pi)}\sin\frac{n\pi}{J+2}\sum_{p=0}^{J-1}\cos
\frac{(2p+3)n\pi}{J+2}\Psi_{(12)}^{p,J-1} \,,
\nonumber
\\
\Xi^{J,n}_{(12)}&=&\tilde\Psi_{(12)}^{J,n}+
\frac{i\,g\sqrt N}{\sqrt 2
(2\pi)}\bigg(-\cos\frac{n\pi}{J+1}{\cal O}_{[14,25]}^{0,J+1}+\sin\frac{n\pi}{J+1}\sum_{p=1}^J \sin \frac{(2p+2)n\pi}{J+2}{\cal
O}_{[14,25]}^{p,J+1}\bigg)\,.\nonumber \\
\end{eqnarray}
As in previous case the coefficient before the ${\cal O}^{0,J+1}$ term
should be considered independently. This term corresponds to the
non-asymptotic term in (3.1) of \cite{Beisert:2005tm}.  The other
operator mixing cases in table \ref{tab:mixing} can be easily obtained
by replacing $i{\cal O}_{[14,25]}^{p,J}$ and $\Psi_{(12)}^{p,J}$ by
the corresponding operators in the same column in table
\ref{tab:mixing}, because they produce the same $H$ matrix.  As we
mentioned in the last section, these operators are not mixed with $D_\mu Z
Z\dots D^\mu Z Z\dots Z$.

\section{Discussion\label{sect:Discuss}}
We have discussed the mixing between single trace BMN operators in the pure
scalar sector and operators with one scalar fermion pair. A kind of
matrix model effective vertices up to $O(g^3)$ are deduced during the
discussion. These effective vertices can also be used to discuss
non-planar diagrams. All the mixing patterns at planar level are
reflected in the vertices (\ref{eq:D-O3}). The $O(g^3)$ matrix
elements in the eigenvalue problem act as a boundary condition to the
$O(g^2)$ matrix.  The results are consistent with
\cite{Georgiou:2008vk} and \cite{Georgiou:2009tp}. It is worth
pointing out that even though the correlation functions between operators
with nonadjacent fermions and pure scalar operators are zero up to
$O(g^3)$ at planar level, they are mixed because the mixings between
operators with adjacent fermions and those with nonadjacent fermions are
needed to diagonalize the one-loop dilatation operator. 

Since we discuss only operators with one fermion pair, the $O(g^2)$
effective vertices do not include the interactions between two fermion
pairs like:
\\
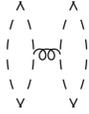
\begin{center}
\begin{picture}(40,40)(10,-20)
\DashCArc(-35,0)(40,-30,30){4}
\DashCArc(35,0)(40,150,210){4}
\DashCArc(-15,0)(40,-30,30){4}
\DashCArc(55,0)(40,150,210){4}
\Gluon(5,0)(15,0){2}{2}
\end{picture}
\end{center}
If one wants to discuss operators with two or more scalar fermion
pairs, one has to include the terms from this diagram which can be
calculated as in \cite{Georgiou:2004ty}.  In general, this term could
also be obtained by using the results in \cite{Beisert:2003jj}. The
$O(g^3)$ terms (\ref{eq:H-O3-Z-A}) and (\ref{eq:H-O3-Z-B}) can be
directly used in more than two fermion pair cases.  It is an
interesting problem to discuss the integrability of this kind of
dynamic spin chain to $O(g^3)$. Unlike the $SU(2|3)$ spin chain in
\cite{Beisert:2003ys,Hernandez:2004kr,Beisert:2005tm}, the fermions
are not invariant under the $SU(3)$ subgroup of $SU(4)_R$. To be
precise, they transform under $SU(2)\times SU(2)$ subgroup.
Understanding how to use the coordinate space Bethe ansatz in this
kind of dynamic spin chain could be a further direction for research. 

One can also discuss the mixing between the $SO(4)$ singlet parts of
$\Psi^{p,J}$ and $V^{p,J}$ along these lines. This requires
calculating the two point correlation function to $O(g^3)$ and we
could expect that the $O(g^3)$ matrix elements act as boundary
conditions, just as in the cases we discussed here. One can also go
further, calculating the dilatation operator matrix between ${\cal
O}^{p,J}$ and $V^{p,J}$ to $O(g^4)$, and we still expect that the
$O(g^4)$ matrix elements act as boundary conditions.  These will
produce the full equations (2.1) and (2.12) in \cite{Georgiou:2009tp}. 

A few remarks on the method in \cite{Georgiou:2008vk}: in
\cite{Georgiou:2008vk}, the authors use the supercurrents deduced from
dimensional reduction of the $D=10$ SUSY YM.  But if one discusses the
quantum properties of ${\cal N}=4$ SUSY YM, one must include the gauge fixing
and ghosts in the quantized Lagrangian.  One would expect that this
should modify the supercurrents, which could be different from those
in \cite{Georgiou:2008vk} in the gauge sector and ghost sector. The
problem is more explicit in ${\cal N}=1$ SUSY YM theory. We can do
gauge fixing and add ghosts in superfield formalism without breaking
${\cal N}=1$ supersymmetry. Then we can have a supercurrent involving
gauge fixing and ghosts. On the other hand, for the original
lagrangian without ghosts and gauge fixing, we also have a
supercurrent. Which one should be chosen to calculate the higher order
correction to the SUSY variation of a composite operator? The method in
\cite{Georgiou:2008vk} is equivalent to choosing the supercurrent
without ghosts. Since all the perturbation calculations are based on
the Lagrangian after quantization, the more consistent choice is to
use the former supercurrent. It turns out that in
\cite{Georgiou:2008vk} when they only discuss the scalar and fermionic
sectors, using those supercurrents from ten-dimensional SUSY YM is
fine.  We can see intuitively why this is not a problem. After careful scrutiny, we can conclude
that their method is consistent using either of these two
currents. This is because after the Faddeev-Popov quantization the
partition function is essentially proportional to the path integral
using the classical action divided by the volume integration of the
gauge redundancy:
\begin{eqnarray}
\langle {\cal O}_1(x_1)\dots {\cal O}_l(x_l)\rangle_{_{\text{F-P}}}=\frac{C\int[\Pi_{n,x} d\phi_n(x)]
{\cal O}_1(x_1)\dots{\cal O}_l(x_l) e^{i S_{c}}}{\int
[\Pi_{\alpha,x}\Lambda^\alpha(x)]\rho[\Lambda(x)]}.
\end{eqnarray}
This is equation (15.5.20) of \cite{Weinberg:1996kr}.  Operators
${\cal O}_i$ are supposed to be gauge invariant. The subscript F-P
means the path integral after Faddeev-Popov quantization and subscript
$c$ means the classical one. $\rho[\Lambda(x)]$ is the invariant
(Haar) measure on the space of group parameters. $\phi_n$ here denotes
generally all the fields to be integrated. Note that the denominator
and the constant $C$ are gauge invariant. We could deduce the Ward
Identity using this path integral with classical action and get back
to Faddeev-Popov ghosts later. Then the current in the Ward Identity
is the classical one without ghosts:
\begin{eqnarray}
&&\hspace{-2cm}
\partial_\mu\langle j_{c}^\mu(x){\cal O}_1(x_1)\dots {\cal
O}_l(x_l)\rangle_{_{\text{F-P}}}
\nonumber
\\&=&\frac{C\int[\Pi_{n,x} d\phi_n(x)]
\partial_\mu j_{c}^\mu(x){\cal O}_1(x_1)\dots{\cal O}_l(x_l)
e^{iS_{c}}}{\int
[\Pi_{\alpha,x}\Lambda^\alpha(x)]\rho[\Lambda]}
\nonumber \\
&=&-\frac{C\int[\Pi_{n,x} d\phi_n(x)]
{\cal O}_1(x_1)\dots{\cal O}_l(x_l)\delta \phi^a(x)\frac{\delta
S}{\delta\phi^a(x)} e^{iS_{c}}}{\int
[\Pi_{\alpha,x}\Lambda^\alpha(x)]\rho[\Lambda]}
\nonumber \\
&=&-i\sum_i\frac{C\int[\Pi_{n,x} d\phi_n(x)]
{\cal O}_1(x_1)\dots\delta{\cal O}_i(x_i)\delta^4(x-x_i)\dots{\cal
O}_l(x_l) e^{iS_{c}}}{\int
[\Pi_{\alpha,x}\Lambda^\alpha(x)]\rho[\Lambda]}
\nonumber
\\&=&-i\sum_i\langle{\cal O}_1(x_1)\dots\delta{\cal
O}_i(x_i)\delta^4(x-x_i)\dots{\cal O}_l(x_l) \rangle_{_{\text{F-P}}}
\label{eq:proof}
\end{eqnarray}
Notice that the infinite integral
in the denominator and numerator in (\ref{eq:proof}) should be understood as
being put on a finite lattice and then taking the infinite limit in
front of the fraction. Since the Faddeev-Popov integral is
well-defined, the limit is well-defined. This can be done in every
line of (\ref{eq:proof}). In their method, the Ward Identity is used
on some operators which are not gauge invariant to deduce the SUSY
transformation property of one operator. Then this SUSY transformation
is used to obtain the SUSY variations of some gauge invariant
operators. This is equivalent to using the Ward Identity on the gauge
invariant operators from the beginning, thus (\ref{eq:proof}) can be
applied.

\section*{Acknowledgements}
I would like to thank George Georgiou, Rodolfo Russo, Valeria Gili and
Tim Morris for helpful discussion and proofreading. I also acknowledge STFC for
financial support. I would also like to thank Gang Yang for passing my
draft to George, Rodolfo and Valeria.

\appendix

\section{$\mathcal{N} =4$ SYM convention\label{sect:Convention}}
The conventions we use in this paper follow closely the paper
\cite{Georgiou:2008vk} except for some normalization differences:
\begin{equation}\label{colcon}
  {\rm Tr}(T^a T^b) \,=\, \delta^{ab}~,~~~
  \left[T^a , T^b \right] \,=\,\sqrt 2\, i f^{abc} T^c~,~~~
  (T^a)^i_j(T^a)^k_l \,=\, 
    \delta^i_l \delta^k_j 
\end{equation}

We choose Minkowski metric $g_{\mu\nu}=diag(+,-,-,-)$. The ${\cal
N}=4$ Yang-Mills Lagrangian is
\begin{multline}
  \label{n4l}
  L  =  {\rm Tr}\left[- \frac{1}{4} F_{\mu \nu} F^{\mu \nu} +
     D_{\mu} {\Phi}_{AB}  D^{\mu} \Phi^{AB}   +  i \psi^{\alpha A}
    \sigma_{\alpha \dot{\alpha}}^{\mu} 
    (D_{\mu} {\bar{\psi}}^{\dot{\alpha}}_{A})
    +\right. \\ 
    \left. \frac 1 2 g^2 [\Phi^{AB}, \Phi^{CD}] [{{\Phi}}_{AB},
    {{\Phi}}_{CD}] -  g  \left([\psi^{\alpha A},
      {{\Phi}}_{AB}]
      \psi_{\alpha}^{B} - [{\bar{\psi}}_{\dot{\alpha} A}, \Phi^{AB}]
      {\bar{\psi}}^{\dot{\alpha}}_{B}\right)\right]
\end{multline}
For scalars, we use $\Phi_{ij}$, $Z_i$, $\bar Z^i$ and $\phi_i$ in
different contexts:
\begin{eqnarray}\label{eq:Z-Phi}
\left\{ \begin{array}{l}
\Phi^{jk}=\frac 1 2 \epsilon^{jki}Z_i
\\
\Phi_{jk}=\frac 1 2 \epsilon_{jki}\bar Z^i
\end{array}
\right.
\hspace{1cm}\left \{
\begin{array}{l}
\Phi^{i4}=\frac 1 2\bar Z^i 
\\
\Phi_{i4}=\frac 1 2  Z_i
\end{array}\right. \hspace{1cm}\text{ for } i,j,k=1,2,3
\end{eqnarray}
where 
\begin{eqnarray}
Z_i=\frac1{\sqrt 2}(\phi_i+i\phi_{i+3})\,, \quad 
\bar Z^i=\frac1{\sqrt 2}(\phi_i-i\phi_{i+3})\,.
\end{eqnarray}
There are some useful relations:
\begin{eqnarray}
\bar Z^{ai} Z^b_j+Z^a_i \bar Z^{bj}
&=&\phi^a_i\phi^b_j+\phi^a_{i+3}\phi^b_{j+3}
\\
Z^{a}_{i} \bar Z^{bj}-\bar Z^{ai} Z^{b}_{j}
&=&-i(\phi^a_i\phi^b_{j+3}-\phi^a_{i+3}\phi^b_{j})
\\
Z^{a}_{i} Z^b_j+\bar Z^{ai} \bar Z^{bj}
&=&\phi^a_i\phi^b_j-\phi^a_{i+3}\phi^b_{j+3}
\\
Z^{a}_{i} Z^b_j-\bar Z^{ai} \bar Z^{bj}
&=&i(\phi^a_i\phi^b_{j+3}+\phi^a_{i+3}\phi^b_{j})
\label{eq:ZtoPhi}
\end{eqnarray}

The propagators are:
\begin{eqnarray}
\langle A^a_\mu(x)A^b_\nu(y) \rangle
&=&-\delta^{ab}g_{\mu\nu}\Delta_{xy}
\,,\nonumber\\
\langle \psi^{a A}_{\alpha}(x)\bar\psi^b_{B\dot{\alpha}}\rangle
&=&i\delta^{ab}\delta^A_B\,
\sigma^\mu_{\alpha\dot{\alpha}}\partial_\mu\Delta_{xy}
\,,\nonumber \\
\langle \Phi^a_{AB}(x)\Phi^b_{CD}(y)\rangle 
&=&\frac 1 4 \delta^{ab}\epsilon_{ { }_{ABCD}}\Delta_{xy}
\,,\nonumber\\
\langle Z^a(x)\bar Z^b(y)\rangle 
&=& \delta^{ab}\Delta_{xy}
\,,\nonumber \\
\langle \bar Z^a(x)\bar Z^b(y)\rangle&=&
\langle Z^a(x) Z^b(y)\rangle =0
\,,\end{eqnarray}
where 
\begin{equation}
\Delta=-\frac 1{4\pi^2(x-y)^2}\,.
\end{equation}

\end{document}